\documentclass[10pt]{article}
\usepackage{graphicx}
\usepackage{amssymb}
\usepackage{epstopdf}
\usepackage{url}

\newcommand{\bibnodot}[1]{}
\title{A Theory of Concepts and Their Combinations I: \vspace{0.3\baselineskip} \\
 \large The Structure of the Sets of Contexts and Properties\footnote{To appear in {\it Kybernetes}, Summer 2004.}}
\author{Diederik Aerts\\
        \normalsize\itshape
        Center Leo Apostel for Interdisciplinary Studies \\
         \normalsize\itshape
         Department of Mathematics and Department of Psychology\\
        \normalsize\itshape
        Vrije Universiteit Brussel, 1160 Brussels, 
       Belgium \\
        \normalsize
        E-Mail: \textsf{diraerts@vub.ac.be} \\ \\
		Liane Gabora \\
		 \normalsize\itshape
		 Center Leo Apostel for Interdisciplinary Studies \\
		  \normalsize\itshape
		  Vrije Universiteit Brussel and Department of Psychology\\
		   \normalsize\itshape
		   University of California, Berkeley, CA 94720-1650, USA \\
		   \normalsize
        E-Mail: \textsf{liane@berkeley.edu}}
\date{}
\begin{document}
\maketitle
\begin{abstract}
\noindent
We propose a theory for modeling concepts that uses the state-context-property theory (SCOP), a generalization of the quantum formalism, whose basic notions are states, contexts and properties. This theory enables us to incorporate context into the mathematical structure used to describe a concept, and thereby model how context influences the typicality of a single exemplar and the applicability of a single property of a concept. We introduce the notion `state of a concept' to account for this contextual influence, and show that the structure of the set of contexts and of the set of properties of a concept is a complete orthocomplemented lattice. The structural study in this article is a preparation for a numerical mathematical theory of concepts that allows the description of the combination of concepts \cite{aertsgabora01}. 
\end{abstract} 

\begin{quotation}
\end{quotation}

\section{Introduction} 
Heinz von Foerster often spoke of how he was introduced to the scientific community in the United States in 1949 \cite{franchietal01}. A prominent role was played by a book he had published in Vienna, which put forward a theory of memory that uses quantum mechanics as an explanatory system \cite{vonfoerster01}. Specifically, he proposed that the forgetting process follows a decay which is the same as radioactive decay, and that remembering happens when one is prompted to retrieve something that has not yet decayed. Excitingly, his `forgetting parameter' turned out to fit perfectly Ebbinghaus' empirical forgetting curve \cite{ebbinghaus01,ebbinghaus02}. Von Foerster also remarked that the decay constants for typical macromolecules---biological molecules---are exactly the same. Thus he suggested that there is a link between the quantum mechanical interpretation of large biological molecules, and our way of keeping things in mind or forgetting them. 

When von Foerster arrived in the United States, Warren McCulloch, head of the Department of Neuropsychiatry at the University of Illinois in Chicago, was intrigued by his ideas about memory. In von Foerster's own words \cite{franchietal01}: ``It turned out that about two or three months before I came there had been a large meeting of some big scientists in America about memory. And they all had lots of data, but no theory. The fascinating thing is that this little booklet of mine had numbers that were matching exactly the data they had \dots. I remember that Warren told me that it was too good to be true, and asked me to give a lecture on the spot. I replied I couldn't give a lecture because of my linguistic inability, but there were so many immigrants---German and Austrian immigrants---at the University of Illinois at that time, that I just had to say it more or less in German and they translated it all very nicely. Then Warren invited me to give another talk. He said ``In two weeks we have a conference in New York; since you're living in New York we invite you," and this was one of the now very famous and legendary Macy meetings." The Macy meetings started the research field of cybernetics, and Heinz von Foerster became one of its leading figures.

When we were invited to contribute to this special issue in honor of von Foerster, it was natural that our contribution would elaborate on his intuition concerning quantum mechanics and the mind. Without knowing of von Foerster's work on it, this had already been one of the research themes in the Center Leo Apostel (CLEA) for some time. In \cite{aertsaerts01} the structure of decision processes in an opinion pole was investigated, and it is shown that the presence of contextual influence gives rise to a nonclassical probability model, specifically one that does not satisfy Kolmogorov's axioms for classical probability theory. Further investigations in this direction yielded a quantum mechanical description of the Liar Paradox \cite{aertsetal05,aertsetal06,aertsetal07}, where it is shown that the contradictory sentences of a multi-sentence Liar Paradox can be represented as an entangled state in a Hilbert space that is the tensor product of Hilbert spaces describing the separate sentences, and the dynamics of the oscillations between truth and falsehood is described by a Schr\"odinger equation. We also distinguished different types of contextuality, investigating the mathematical structure they give rise to, and showed that the kind of contextual interaction that arises in quantum mechanics, and thus the mathematical structure necessary to describe this contextual interaction, also appears in cognition \cite{aertsetal01,aertsetal04}. This led to the development of a {\it contextualized} theory of concepts, the mental sieves through which memories are categorized, organized, and creatively blended to make sense of experiences \cite{gabora01,gaboraaerts01,gaboraaerts02}.

Concepts are what we use to navigate through and make sense of the world around us, enabling us to classify and interpret new situations in terms of previous similar ones. They can be concrete, like `chair', or abstract, like `beauty'. A concept is generally associated with a set of {\it properties}. For example, the concept `chair' is associated with the property 'has four legs'. Something that is a property of a particular concept can also be a concept itself. Thus `scaly' is not just a property of `fish', but a concept in its own right. 

According to the {\it classical} or rule-based view of concepts, which goes back to Aristotle, all instances of a concept share a common set of necessary and sufficient defining properties. Wittgenstein pointed out that it is not possible to give a set of characteristics or rules that define a concept. For example, how would one define `game' such that frisbee, baseball, and roulette are classified as games, while wars, debates, and leisure walking are not \cite{wittgenstein01}?  Furthermore, it is often unclear whether an object is a member of a particular category; {\it e.g.} whether a camel can be considered a vehicle \cite{goldstonekersten01}. One might hypothesize that this ambiguity stems from individual differences in categorization rules. But in \cite{mccloskeyglucksberg01} it is shown that subjects will categorize an object as a member of one category at one time and a member of another category at a different time. Other problems with the classical view of concepts are reviewed in \cite{smithmedin01} and \cite{komatsu01}.  

The critical blow to the classical view came from work on color; it was shown that colors do not have any particular criterial attributes or definite boundaries, and instances differ with respect to how typical or exemplary they are of a category \cite{rosch05}. This led to formulation of the {\it prototype theory}  \cite{roschmervis01,rosch03,rosch04}, according to which concepts are organized around family resemblances, and consist of not defining, but {\it characteristic} features, which are weighted in the definition of the prototype. Rosch showed that subjects rate concept membership as graded, with degree of membership of an instance corresponding to conceptual distance from the prototype. Moreover, the prototype appears to be particularly resistant to forgetting \cite{homaetal01}. The prototype theory also has the strength that it can be mathematically formulated and empirically tested. By calculating the similarity between the prototype for a concept, and a possible instance of it, across all salient features, one arrives at a measure of {\it conceptual distance} between the instance and the prototype. Another means of calculating conceptual distance comes out of the {\it exemplar theory}, \cite{nosofsky02,nosofsky03,medinetal01,heitbarselou01} according to which a concept is represented by, not a set of defining or characteristic features, but a set of salient {\it instances} of it stored in memory. The exemplar model has met with considerable success at predicting results \cite{nosofsky03,tenpenny01}. Moreover, there is indeed evidence of preservation of specific training exemplars in memory \cite{thomas01}. Although prototype and exemplar theories have been extensively pitted against one another, neither 
cannot fully reproduce individual differences in the distributions of responses across test stimuli \cite{nosofskyetal01}, or account for certain base-rate effects in categorization \cite{nosofskyetal02}. Classical, prototype, and exemplar theories are sometimes referred to as `similarity based' approaches, because they assume that categorization relies on data-driven statistical evidence. They have been contrasted with `explanation based' approaches, according to which categorization relies on a rich body of knowledge about the world \cite{goldstonekersten01}. For example, according to the {\it theory theory} concepts take the form of `mini-theories' \cite{murphymedin01} or schemata \cite{rumelhartnorman01}, in which the causal relationships amongst properties are identified.

None of the existing theories on concepts describes `how concepts combine', {\it i.e.} derive the model that represents the combination of two or more concepts from the models that represent the individual concepts. The {\it combination problem} is considered so serious
that it has been said that not much progress is possible in the field if no light is shed on this problem \cite{fodor01,kamppartee01,rips03,hampton01}. Directly related to the combination problem, already in the eighties, the so-called `guppy effect' was identified, where {\it guppy} is not rated as a good example
of `pet', nor of `fish', but it is rated as a good example of the combination `pet-fish' \cite{oshersonsmith01}. General fuzzy set theory \cite{zadeh01} has been tried in vain to deliver a description of the guppy effect \cite{zadeh02,oshersonsmith02}, and also intuitively it is possible to understand the peculiarity: if (1) activation
of `pet' causes a small activation of {\it guppy}, and (2) activation of `fish' causes
a small activation of {\it guppy}, how is it that (3) activation of `pet-fish' causes
a large activation of {\it guppy}? Also the explanation based theories, since they have not lent themselves to mathematical formulation, have not been able to model what happens when concepts combine \cite{komatsu01,fodor01,rips03}. 

In \cite{aertsgabora01} we show explicitly that our theory of concepts in Hilbert space can model an arbitrary combination of
concepts by making use of the standard quantum mechanical procedure to describe the combinations of quantum entities. We show that also the guppy effect is modeled in a natural way by our theory. In the present article we prepare the Hilbert space description that is elaborated in \cite{aertsgabora01}.

\section{The State Context Property Formalism} \label{sec:scop}
The basic ingredients of our theory are `states', `contexts' and `properties', and thus the models built are referred to as State Context
Property Systems (SCOPs). In this section we introduce the basics of our formalism.

\subsection{Contexts, States and Properties}
Although traditionally the main function of concepts was to represent a class of entities in the world, increasingly they are thought to have no fixed representational structure, their structure being spontaneously evoked by the situations in which they arise \cite{riegleretal01,rosch06}.

Rosch's insight, and the basis for the `similarity based theories', was that the typicality of different exemplars and the applicability of different properties of one and the same concept vary. As such, subjects rate different typicalities for exemplars of the concept `fruit', {\it e.g.} delivering
the following classification with decreasing typicality: {\it apple, strawberry, plum,
pineapple, fig, olive}. For the concept `sports' the exemplars {\it
football, hockey, gymnastics, wrestling, archery, weightlifting} are classified
for decreasing typicality, and for the concept `vegetable'
this happens with {\it carrot, celery, asparagus, onion, pickle, parsley} \cite{rosch02,rosch01,armstrongetal01}. The insight
of our theory is that `for each exemplar alone' the typicality varies with respect to the context that influences it. In an analogous way `for each
property alone', the applicability varies with respect to the context. We performed
an experiment (Section \ref{sec:experiment}) where this `contextual typicality and applicability effect' is shown and measured.
Subjects classify exemplars of the concept `pet' under different contexts, {\it e.g.} the
context, `The pet is chewing a bone', which results in a classification with decreasing
typicality as follows: {\it dog, cat, rabbit, hamster, guinea pig, mouse,
hedgehog, bird, parrot, snake, canary, goldfish, spider, guppy} (Table \ref{tab:ratingexemplars}). The same exemplars
are classified differently in decreasing typicality for the context, `The pet is being
taught': {\it dog, parrot, cat, bird, hamster, canary, guinea pig, rabbit,
mouse, hedgehog, snake, goldfish, guppy, spider}, and again differently for the
context, `Did you see the type of pet he has? This explains that he is a weird person': {\it spider, snake,
hedgehog, mouse, rabbit, guinea pig, hamster, parrot, bird, cat, dog, canary,
goldfish, guppy} (Table \ref{tab:ratingexemplars}). The effect is also measured for the applicability of a property (Table \ref{tab:applicabilityratings}). 

This `contextual typicality and applicability effect' can be described by introducing the
notion of {\it state of a concept}, and hence consider a concept an entity that can be in
different states, and such that a context provokes a change of state of the concept.
Concretely, the concept `pet' is in another state under the context, `The pet is chewing a
bone' than under the context, `Did you see the type of pet he has? This explains that he is a weird person'.
It is the set of these states and the dynamics of change of state under the influence of
context that is modeled by SCOP and by our quantum mechanical
formalism in Hilbert space. The problem of the combination of concepts gets resolved in
our theory because in combination, the concepts are in different states; for example in
the combination `pet-fish', the concept `pet' is in a state under the context `the pet is
a fish', while the concept `fish' is in a state under the context `the fish is a pet'.
The states of `pet' and `fish' under these contexts have different typicalities, which explains the guppy effect. Hence, a {\it context} is a `relevant context' for a concept if it changes the state of the concept which manifests experimentally as a change in the typicality of exemplars and the applicability of properties. A context can itself be a concept, or aggregation of concepts, or it can be a goal or drive state, a previous lingering thought, feeling, or experience, or ones' physical surrounding. Since in this article we focus on the description of the combination of concepts the contexts that we consider are aggregations of concepts, because it are this type of contexts that play a role in way concepts combine.

The set of relevant properties of a concept is in SCOP such that also less characteristic properties are included. Let us explain why this is he case. If the focus of a theory of concepts is on its nature as an identifier the tendency is to concentrate on the most characteristic aspects of the concept (be they typical properties, salient exemplars, or characteristic instances), which means that less characteristic or contextually-determined properties will be excluded. The danger of this becomes evident when we look again at the `pet fish' example. The presence of the guppy effect shows that something happens to the concept `pet' when it combines with the concept `fish', and vice versa; they interact. Moreover, it shows they interact in a way that would be hard to predict; one might expect that if an instance is typical of one concept and typical of another, it would be particularly typical of their combination, but the opposite is true here. The most characteristic properties---those generally included in a model of categorization---are barely influenced in interactions with other concepts. Whenever the concept appears, the property is present, and to more or less the same degree. For example, it is not via the property `lives in a house', typical of pets, that we can detect that the concept `pet' is `interacting' with the concept `fish'. It is via a less characteristic property of pets such as `likes to sit on its owner's lap', which is lost completely when `pet' interacts with `fish' in the combination `pet fish'. A model that contains enough fine structure to model interactions must incorporate those properties that are sensitive to interactions and subject to change. Hence, apart from states, that describe the `contextual typicality and applicability effect' that we mentioned, our theory does not focus alone on the most characteristic properties but also introduces context sensitive and less characteristic properties.

Before we elaborate the SCOP theory we fix some notations. If we make claims about an arbitrary concept, it will be denoted $S$ and its set of states $\Sigma$ (or $\Sigma^S$ and $\Sigma^T$ if more than one concept---for example two concepts $S$ and $T$---are considered). Individual states of
concept $S$ will be denoted $p, q, r, \ldots \in \Sigma$. We introduce one special state of a concept $S$ called the {\it ground state}, denoted $\hat{p}$.
One can think of the ground state as the state the concept is in when it is not triggered by any particular context. We explain in Section 4.6 of \cite{aertsgabora01} that it is the state a concept is in when it constitutes a sub-concept of the compound of all concepts held in the mind. The set of contexts relevant to a concept $S$ will be denoted ${\cal M}$ (or ${\cal M}^S$ and ${\cal M}^T$ if more than one concept, for example two concepts $S$ and $T$, are considered), and individual contexts by $e, f, g, \ldots
\in {\cal M}$, and the set of properties of the concept $S$ will be denoted by ${\cal L}$ (or ${\cal L}^S$ and ${\cal L}^T$ if more than one concept, for example two concepts $S$ and $T$, are considered), and individual properties by $a, b, c \ldots \in {\cal L}$. Consider the concept `pet' in its ground state $\hat{p}$ and consider the context
\begin{equation}
e_1: {\rm `The\ pet\ is\ chewing\ a\ bone'}
\end{equation}
This context consists of the situation where `a pet is chewing a bone'. If an example involves several states, contexts and
properties, we use subscripts. Thus $p_1, p_2, p_3,
\ldots, p_n,
\ldots
\in
\Sigma$ denote states, 
$e_1, e_2, e_3, \ldots, e_m,
\ldots \in {\cal M}$ denote contexts, and $a_1, a_2, a_3, \ldots, a_k, \ldots \in {\cal L}$ denote properties. The context $e_1$ for the concept `pet' will cause the ground state
$\hat{p}$ of `pet' to change to another state, $p_1$. That $\hat{p}$ and $p_1$ are different states is manifested by fact that the frequency measures of different exemplars of the concept will be different for different states (Table \ref{tab:ratingexemplars}), and the properties will have different applicability values for different states (Table \ref{tab:applicabilityratings}). In relation with Table \ref{tab:ratingexemplars} we remark that rather than typicality values of an exemplar under different contexts we need frequency values of this exemplar under different contexts for our theory. Typicality and frequency are linked, in the sense that an increasing frequency will generate an increasing typicality, and vice versa. But typicality contains more aspects than just frequency. It is well
possible that for two exemplars with equal frequency, the typicality of one of them is higher than for the other, due to the fact that
although both exemplars are equally abundant in the considered context, one of them is still more typical than the other one. The aspects
contained in typicality that are extra to frequency will be described in our theory in a different way than the aspects related to
frequency.

\subsection{The Experiment} \label{sec:experiment}
The `contextual frequency and applicability effect' that we mentioned in the foregoing section was tested in the following experiment. The 81 participating subjects---email correspondents, {\it i.e.} friends and colleagues of the experimenters---were presented a questionnaire by e-mail as attached file, which could then be filled out and sent back. The questionnaire was accompanied by the following text:

\medskip
\noindent 
{\small {\it This study has to do with what we have in mind when we use words that refer to categories. Think of the category `fruit' and the contextual situation expressed by the sentence: ``The fruit gets squeezed for a fresh drink of juice''. The examples ``orange" and ``lemon" {\em appear more frequently} in this contextual situation than do the examples ``strawberry" or ``apple" and certainly than ``fig" or ``olive".

It is the frequency with which examples of a category appear in a specific contextual situation that we want you to estimate in this experiment. You are to rate this frequency for each example on a 7-point scale. When you fill out {\bf 7}, this means that you feel that the example appears very frequently in the given contextual situation. A {\bf 1} means you feel the example appears very rarely and a {\bf 0} means not at all. A {\bf 4} means that you feel the example appears moderately frequently. Use the intermediate numbers {\bf 2}, {\bf 3}, {\bf 5}, and {\bf 6} to express intermediate judgments.

You will see that each test consists of an A part, where we test the frequency that an example of a category appears in a specific situation, and a B part. In this B part we test the rate of applicability of specific properties to the category in the given situation. Consider again the category `fruit' and the contextual situation ``The fruit gets squeezed for a fresh drink of juice". The properties ``mellow" and ``tasty" {\em are more applicable} to `fruit' in this contextual situation than the properties ``unripe" and ``moldy". Also the applicability is rated on a 7-point scale, {\bf 7} meaning very applicable, {\bf 1} meaning almost not applicable, with {\bf 0}, not applicable at all, and {\bf 4} meaning moderately applicable. Use again the intermediate numbers {\bf 2}, {\bf 3}, {\bf 5}, and {\bf 6} to express intermediate judgments.

When you estimate the frequency of a specific example in a situation this can refer to the amount of times that you personally have experienced this example in this context. But your estimation of this frequency may also relate to confrontations with this example on TV, in movies, in dreams, in your imagination, {\it etc} \ldots. Don't worry about why you estimate that way for a certain example or property. And don't worry about whether it's just you or people in general who estimate that way. Just mark it the way you feel it.}}

\medskip
\noindent
Let us examine the results obtained when subjects were asked to rate the frequency with which a particular exemplar might be encountered in a specific context for the concept `pet'. The contexts are given in Table \ref{tab:differentcontexts}, and are of the type `congregations of concepts'.
\begin{table}[h] 
\small \begin{center}
\begin{tabular}{|l|l|}
\hline
$e_1$ & The pet is chewing a bone \\
\hline
$e_2$ & The pet is being taught \\
\hline
$e_3$ & The pet runs through the garden \\ 
\hline
$e_4$ & Did you see the type of pet he has? This explains that he is a weird person \\
\hline
$e_5$ & The pet is being taught to talk \\
\hline
$e_6$ & The pet is a fish \\
\hline
$1$ & The pet is just a pet (the unit context for `pet')\\
\hline
\end{tabular}
\end{center}
\caption{The contexts considered in the experiment.} \label{tab:differentcontexts}
\end{table}
Subjects were presented seven different contexts, here called $e_1, e_2, \ldots, e_6, 1$, and asked to rate on a scale between 1 and 7, the frequency
with which a specific exemplar of the concept `pet' appears in this context. The 14 exemplars proposed were: {\it rabbit, cat, mouse, bird, parrot, goldfish, hamster, canary, guppy, snake, spider, dog, hedgehog}, and {\it guinea pig}. The results of the ratings are given in Table
\ref{tab:ratingexemplars}, for each exemplar and context under `rate', while the relative frequency, calculated from this rating, is given for each
exemplar and context under `freq.'. 
\begin{table}[h] 
\small
\begin{center}
\begin{tabular}{|l||l|l|l|l|l|l|l|l|l|l|l|l|l|l|} \hline 
\multicolumn{1}{|c||}{exemplar} &  \multicolumn{2}{|c|}{$e_1$} & \multicolumn{2}{|c|}{$e_2$} &
\multicolumn{2}{|c|}{$e_3$} &
\multicolumn{2}{|c|}{$e_4$} & \multicolumn{2}{|c|}{$e_5$} & \multicolumn{2}{|c|}{$e_6$} & \multicolumn{2}{|c|}{$1$}\\
\hline   
& rate & freq  & rate & freq & rate & freq & rate & freq & rate & freq & rate & freq & rate & freq \\ \hline 
{\it rabbit}  & 0.07  & 0.04 & 2.52 & 0.07 & 4.58    & 0.15 & 1.77 & 0.05 & 
 0.15  & 0.01 & 0.10 & 0.00 &  4.23 & 0.07 \\ \hline
{\it cat}  & 3.96  & 0.25 & 4.80  & 0.13 & 6.27    & 0.22 &  0.94  & 0.03 &  0.46  & 0.03 &  0.15 & 0.01 & 6.51 & 0.12 \\ \hline

{\it mouse}  &  0.74  & 0.03 & 2.27   & 0.06 &  2.67   & 0.08 &  3.31 & 0.11 & 0.12 & 0.01 & 0.10 & 0.00 &  2.59 & 0.05 \\ \hline
{\it bird} & 0.42  & 0.02 & 3.06   & 0.08 &  0.63   & 0.02 &  1.41 & 0.04 & 
 2.21 & 0.17 & 0.15 & 0.01 & 4.21 & 0.08 \\ \hline
{\it parrot}  & 0.53  & 0.02 &  5.80   & 0.16 &  0.44   & 0.01 &  1.57  & 0.04 & 6.72 & 0.63 & 0.16 & 0.01 & 4.20 & 0.07 \\ \hline
{\it goldfish}  & 0.12  & 0.01 &  0.69   &  0.02 & 0.09   & 0.00 & 0.83 & 0.02 & 0.10 & 0.00 & 6.84 & 0.48 & 5.41 & 0.10 \\ \hline
{\it hamster}  & 0.85  & 0.04 & 2.72    & 0.07 & 2.06   & 0.06 & 1.25  & 0.04 & 0.14 & 0.01 & 0.09 & 0.00 & 4.25 & 0.07 \\ \hline
{\it canary}  &  0.26  & 0.01 & 2.73   & 0.07 & 0.23   & 0.01 &  0.86  & 0.02 & 1.08 & 0.07 & 0.14 & 0.01 & 4.79 & 0.08 \\ \hline
{\it guppy}  & 0.14   & 0.01 & 0.68   & 0.02 & 0.09    & 0.00 & 0.83 & 0.02 & 
 0.10  & 0.00 &  6.64 & 0.46 & 5.16 & 0.09 \\ \hline
{\it snake}  & 0.57  & 0.02 &  0.98    & 0.02 & 0.36   & 0.01 & 5.64  &  0.22 & 0.09 & 0.00 & 0.15 & 0.01 & 1.60 & 0.03 \\ \hline
{\it spider}  & 0.26  & 0.01 & 0.40  & 0.01 & 1.05    & 0.03 & 5.96  & 0.23 & 
 0.09 & 0.00 & 0.09 & 0.00 &  1.22 & 0.02 \\ \hline
{\it dog}  & 6.81  & 0.50 & 6.78   & 0.19 & 6.85    & 0.24 & 0.91 & 0.03 &  
 1.02 & 0.06 & 0.11 & 0.00 & 6.65 & 0.12 \\ \hline
{\it hedgehog}  & 0.53  & 0.02 & 0.85  & 0.02 & 2.59     & 0.08 & 3.48  & 0.12 & 0.11 & 0.00 & 0.09 & 0.00 & 1.56 & 0.03 \\ \hline
{\it guinea pig}  & 0.58  & 0.03 & 2.63   & 0.07 & 2.79   & 0.09 & 1.31 & 0.04 & 0.15 & 0.01 & 0.09 & 0.00 & 3.90 & 0.07 \\ \hline
\end{tabular}
\end{center}
\caption{Frequency ratings of different exemplars for different contexts.} \label{tab:ratingexemplars}
\end{table}
As an example of how to interpret these data, consider the context $e_1$, `The pet is chewing a bone'. Of 100 situations of this kind, the subjects estimated that this pet would be a {\it rabbit} in 4, a {\it cat} in 25, a {\it mouse} in 3, a {\it bird} in 2, a {\it parrot} in 2, a {\it goldfish} in 1, a {\it hamster} in 4, a {\it canary} in 1, a {\it guppy} in 1, a {\it snake} in 2, a {\it spider} in 1, a {\it dog} in 50, a {\it hedgehog} in 2, and a {\it guinea pig} in 3 situations (Table \ref{tab:ratingexemplars}). The context $1$ is the {\it unit context}. Here the subject had to estimate the frequency of the different exemplars of the concept 'pet' in the presence of any arbitrary context; hence in the absence of a specific context. This means that the frequencies retrieved with context $1$ correspond to the frequencies represented by the ground state $\hat{p}$ of the concept `pet'. By means of this experiment we are now able to explain some of the more subtle aspects of the proposed formalism. The ground state has to describe the frequencies retrieved by means of the unit context $1$, hence in 100 situations of pets in any context, subjects estimated that there are 7 situations where the pet is a {\it rabbit}, 12 where it is a {\it cat}, 5 where it is a {\it mouse}, 8 where it is a {\it bird}, 7 where it is a {\it parrot}, 10 where it is {\it goldfish}, 7 where it is a {\it hamster}, 8 where it is a {\it canary}, 9 where it is a {\it guppy}, 3 where it is a {\it snake}, 2 where it is a {\it spider}, 12 where it is a {\it dog}, 3 where it is a {\it hedgehog}, and 7 where it is a {\it guinea pig} (Table \ref{tab:ratingexemplars}). Each of the considered contexts gives rise to another state of the concept `pet'. Let us call $p_1, p_2, p_3, \ldots, p_6$ the states obtained after a change provoked by contexts $e_1, e_2, e_3, \ldots, e_6$ on the ground state $\hat{p}$.

We have also tested the applicability values of different properties of pet in different contexts, using the 14 properties in Table \ref{tab:properties}. 
\begin{table}[h] 
\small \begin{center}
\begin{tabular}{|l|l|}
\hline
$a_1$ & {\it lives in and around the house} \\
\hline
$a_2$ & {\it furry} \\
\hline
$a_3$ & {\it feathered} \\ 
\hline
$a_4$ & {\it likes to sit on owners lap} \\
\hline
$a_5$ & {\it likes to be caressed} \\
\hline
$a_6$ & {\it can fly} \\
\hline
$a_7$ & {\it can swim} \\
\hline
$a_8$ & {\it likes to swim} \\
\hline
$a_9$ & {\it noisy} \\
\hline
$a_{10}$ & {\it silent} \\
\hline
$a_{11}$ & {\it scary} \\
\hline
$a_{12}$ & {\it hairy} \\
\hline
$a_{13}$ & {\it docile} \\
\hline
$a_{14}$ & {\it friendly} \\
\hline
\end{tabular}
\end{center}
\caption{The properties of `pet' considered in the experiment.} \label{tab:properties}
\end{table}
Subjects were asked to rate the applicability of each of the properties in Table \ref{tab:properties} for the same seven contexts. The results are presented in Table \ref{tab:applicabilityratings}. Under `rate' are the ratings on a 7-point scale, and under `wt' is the renormalization to a number between 0 and 1, called the weight of the property.
\begin{table}[h] 
\small
\begin{center}
\begin{tabular}{|l||l|l|l|l|l|l|l|l|l|l|l|l|l|l|l|l|l|l|l|l|l|l|l|l|l|l|l|} \hline 
\multicolumn{1}{|c||}{} & \multicolumn{2}{|c|}{$e_1$} & \multicolumn{2}{|c|}{$e_2$} & \multicolumn{2}{|c|}{$e_3$} &
\multicolumn{2}{|c|}{$e_4$} & \multicolumn{2}{|c|}{$e_5$} & \multicolumn{2}{|c|}{$e_6$} & \multicolumn{2}{|c|}{$1$}\\
\hline   
& rate & wt  & rate & wt & rate & wt & rate & wt & rate & wt & rate & wt & rate & wt \\ \hline 
$a_1$ &5.99 & 0.86 & 6.11  & 0.87 & 6.51  & 0.92 &  2.95  & 0.42 & 4.5 & 0.64 & 4.09 & 0.58 & 6.60 & 0.94 \\ \hline
$a_2$ & 4.65  & 0.66 & 4.09   & 0.58 & 4.99  & 0.71 &  1.64 & 0.23 & 1.65 & 0.23 & 0.30 & 0.04 & 4.83 & 0.69 \\ \hline
$a_3$ & 0.58  & 0.08 & 2.95   & 0.42 & 1.15  & 0.16 & 1.73 & 0.25 & 5.86 &  0.84 & 0.09  & 0.01 & 4.11 & 0.59 \\ \hline
$a_4$ & 4.79  & 0.68 & 5.41   & 0.77 &  5.09  & 0.73 & 1.27   & 0.18 & 1.80 & 0.26 & 0.09 & 0.01 & 5.32 & 0.76 \\ \hline
$a_5$ &  5.06  & 0.72 & 5.53    & 0.79 &  5.19  & 0.74 & 1.40  & 0.20 & 2.51 & 0.36 & 0.43 & 0.06 & 5.37 & 0.77 \\ \hline
$a_6$ & 0.44  &  0.06 & 2.65  & 0.38 &  0.81  & 0.12 & 1.72 & 0.25 & 5.72 & 0.82 & 0.20 & 0.03 & 3.65 & 0.52 \\ \hline
$a_7$ & 3.44  & 0.49 & 2.91    & 0.42 & 2.83  &  0.40 & 1.79 & 0.26 & 0.84 & 0.12 & 6.99 & 1.00 & 4.23 & 0.60 \\ \hline
$a_8$ & 3.36  & 0.48 & 3   & 0.43 & 2.99  & 0.43 & 1.48  & 0.21 & 0.70 & 0.10 & 6.43 & 0.92 & 3.86 & 0.55 \\ \hline
$a_9$ & 4.33  & 0.62 & 3.42   & 0.49 & 4.05  & 0.58 & 2.80  & 0.40 & 4.81 & 0.69 & 0.20 &  0.03 & 3.68 & 0.53 \\ \hline
$a_{10}$ & 2.70  & 0.39 & 2.83   & 0.40 & 2.73  & 0.39 & 3.14   & 0.45 & 1.52 & 0.22 & 5.99 & 0.86 & 3.46 & 0.49 \\ \hline
$a_{11}$ & 2.74  & 0.39 & 1.33   & 0.19 & 1.99  & 0.28 & 5.93  & 0.85 & 0.79 & 0.11 & 1.20 & 0.17 &  1.30 & 0.18 \\ \hline
$a_{12}$ & 4.63  & 0.66 & 3.75   & 0.54 & 4.62  & 0.66 & 3.07   & 0.44 & 0.96 & 0.14 & 0.12 & 0.02 &  4.28 & 0.61 \\ \hline
$a_{13}$ & 4.49  & 0.64 & 5.17   & 0.74 & 4.74  & 0.68 & 1.25  & 0.18 &  3.43 & 0.49 & 3.48 &  0.50 & 5.20 & 0.74 \\ \hline
$a_{14}$ & 4.40  & 0.63 & 5.19   & 0.74 & 4.88  & 0.70 & 1.15  & 0.16 &  3.40 & 0.49 & 2.11 & 0.30 & 5.01 & 0.72 \\ \hline
\end{tabular}
\caption{Applicability of properties in different contexts.} \label{tab:applicabilityratings}
\end{center}
\end{table}
Consider `pet' under the context $e_1$, `The pet is chewing a bone'. We see that property $a_2$, {\it furry} has a weight of 0.66, while
under the context $e_4$, `Did you see the type of pet he has? This explains that he is a weird person' it has a much lower weight of 0.23. On the
other hand, property $a_3$, {\it feathered} has a weight of 0.08 under the context $e_1$, `The pet is chewing a bone', which is extremely low, while it has weight
0.84 under the context $e_5$, `The pet is being taught to talk'.
 
We performed a statistical analysis of the data, using the `t-test for paired two samples for means' to estimate the probability that the shifts of means under the different contexts is due to chance. The full analysis including histograms can be obtained by contacting the authors. In all but a few cases, the effect of context was highly significant. Let us give some concrete examples, and also comment on the few exceptions.

As shown in Table \ref{tab:ratingexemplars}, the mean frequency of {\it rabbit} under context $e_1$ `The pet is chewing a bone', is 0.04, while under context $e_2$ `The pet is being taught', it is 0.07. The p-value for {\it rabbit} for the two contexts $e_1$ and $e_2$ is $6.79 \times 10^{-5}$; thus we can strongly reject the null hypothesis that the two means are identical and that the measured difference in mean is due to chance. Hence the measured difference in mean reflects a genuine context effect. In Table \ref{tab:pvaluesexemplars} p-values for the other exemplars are presented. The means under contexts $e_1$ and $e_2$ for {\it mouse, bird, parrot, goldfish, hamster, canary, guppy} and {\it dog} are sufficiently different that the p-values are small enough to reject the null hypothesis of no difference (Table \ref{tab:ratingexemplars}). This is not the case for {\it snake, spider} and {\it hedgehog}, which were estimated with very low frequency for both $e_1$ and $e_2$. The resulting high p-value indicates that the difference in mean could be due to statistical fluctuations, hence a genuine effect of context. Table \ref{tab:pvaluesexemplars} shows that the exemplars most relevant to the distinction between contexts $e_1$, `The pet is chewing a bone', and $e_2$, `The pet is being taught', have the lowest p-value. {\it Parrot} has the lowest p-value of all, namely 6.30E-28. Indeed, parrots never chew bones, while they can certainly be taught. Then come {\it bird} with 9.16E-20, {\it canary} with 3.17E-21 and {\it dog} with 2.34E-21. For {\it bird} and {\it canary} a similar observation can be made as for {\it parrot}; they have little affinity with `chewing a bone', while quite some with `being taught' (but less than {\it parrot}). For {\it dog}, the subjects rated the affinity with `chewing a bone' much higher than for `being taught', thus the small p-value. Next come {\it cat} and {\it guinea pig} with very small p-values of 6.00E-9 and 1.53E-9. Subjects rated {\it cat} with more affinity for `chewing a bone' and less for `being taught', and {\it vice versa} for {\it guinea pig}. Then come {\it rabbit, mouse, goldfish, hamster}, and {\it guppy}, with very small p-values. Although they are definitely less significant with respect to contexts $e_1$ and $e_2$, the subjects rated them with significantly less affinity for `chewing a bone' than for `being taught'. Thus they have much smaller p-values than {\it snake, spider}, and {\it hedgehog}, which subjects rated equally irrelevant for distinctions involving `chewing a bone' and `being taught'.
\begin{table}[h] 
\scriptsize
\begin{center}
\begin{tabular}{|l||l|l|l|l|l|l|l|} \hline 
exemplar & p-value & p-value & p-value & p-value & p-value & p-value & p-value \\ \hline 
& $e_1/e_2$ & $e_1/e_3$ & $e_1/e_4$ & $e_1/e_5$ & $e_1/e_6$ & $e_1/1$ & $e_2/e_3$ \\ \hline
{\it rabbit}  & 6.79E-5  & 1.85E-21 & 0.053 & 2.27E-5 & 4.87E-6    & 3.44E-7 & 5.14E-19  \\ \hline
{\it cat}  & 6.00E-9  & 0.15 & 3.03E-20  & 3.65E-20 & 1.83E-22    & 2.48E-10 &  3.24E-25   \\ \hline
{\it mouse}  &  1.17E-5  & 1.33E-8 & 7.98E-18   & 1.99E-6 &  9.55E-7   & 0.009 &  0.017  \\ \hline
{\it bird} & 9.16E-20  & 0.94 & 0.13E-3   & 3.20E-15 &  0.77E-3   & 2.69E-22 &  7.12E-18  \\ \hline
{\it parrot}  & 6.30E-28  & 0.07 &  0.012   & 1.23E-34 &  0.013   & 5.73E-10 &  4.78E-45   \\ \hline
{\it goldfish}  & 4.27E-5  & 0.0015 &  6.84E-8   &  0.47 & 3.34E-59   & 3.56E-45 & 1.72E-7  \\ \hline
{\it hamster}  & 4.42E-5  & 1.33E-6 & 0.71    & 7.89E-6 & 2.02E-6   & 9.20E-6 & 0.16   \\ \hline
{\it canary}  &  3.17E-21  & 1.97E-15 & 0.12E-3   & 2. 75E-7 & 0.47   & 7.80E-38 &  8.11E-22   \\ \hline
{\it guppy}  & 1.36E-3  & 0.73 & 3.35E-5   & 0.44 & 1.56E-54    & 9.02E-35 & 1.53E-8  \\ \hline
{\it snake}  & 0.96  & 0.52E-3 &  2. 38E-23    & 0.92E-3 & 0.46E-2   & 0.72 & 7.23E-5   \\ \hline
{\it spider}  & 0.80  & 0.94 & 1.02E-29  & 0.71E-1 & 0.59E-1    & 0.12E-1 & 5.96   \\ \hline
{\it dog}  & 2.34E-21  & 7.21E-31 & 7.18E-31   & 3.34E-25 & 1.18E-32    & 3.03E-26 & 3.43E-7  \\ \hline
{\it hedgehog}  & 0.90  & 2.13E-32 & 6.59E-17  & 0.16E-3 & 7.20E-5     & 0.14 & 3.48   \\ \hline
{\it guinea pig}  & 1.53E-9  & 1.28E-9 & 0.13   & 7.57E-5 & 1.15E-5   & 1.60E-10 & 1.31  \\ \hline
 &  $e_2$/$e_4$ & $e_2$/$e_5$ & $e_2$/$e_6$ & $e_2$/$1$ & $e_3$/$e_4$ & $e_3$/$e_5$ & $e_3$/$e_6$\\ \hline   
{\it rabbit}  & 0.69E-1  & 5.25E-20 & 2.50E-22 & 2.27E-5 & 4.87E-6    & 3.44E-7 & 5.14E-19  \\ \hline
{\it cat}  & 3.59E-29  & 5.91E-27 & 1.19E-40  & 3. 65E-20 & 1.83E-22    & 2.48E-10 &  3.24E-25   \\ \hline
{\it mouse}  &  4.23E-7  & 2.38E-19 & 1.65E-20   & 1.99E-6 &  9.55E-7   & 0.90E-2 &  0.17E-1  \\ \hline
{\it bird} & 9.15E-8  & 1.03E-6 & 4.99E-25   & 3.20E-15 &  0.77E-3   & 2.69E-22 &  7.12E-18  \\ \hline
{\it parrot}  & 4.64E-28  & 7.14E-29 &  8.03E-46   & 1.23E-34 &  0.13E-1   & 5.73E-10 &  4.78E-45   \\ \hline
{\it goldfish}  & 0.14  & 8.31E-6 &  4.67E-60   &  0.47 & 3.34E-59   & 3.56E-45 & 1.72E-7  \\ \hline
{\it hamster}  & 5.01E-9  & 7.65E-25 & 7.94E-27    & 7.89E-6 & 2.02E-6   & 9.20E-6 & 0.16   \\ \hline
{\it canary}  &  4.64E-14  & 0.52 & 4.61E-19   & 2. 75E-7 & 0.47   & 7.80E-38 &  8.11E-22   \\ \hline
{\it guppy}  & 0.09  & 1.81E-6 & 1.89E-54   & 0.44 & 1.56E-54    & 9.02E-35 & 1.53E-8  \\ \hline
{\it snake}  & 4.31E-27  & 1.29E-8 &  1.25E-5    & 0.92E-3 & 0.46E-2   & 0.72 & 7.23E-5   \\ \hline
{\it spider}  & 2.54E-32  & 0.11E-2 & 0.45E-3  & 0.71E-1 & 0.59E-1    & 0.12E-1 & 5.96   \\ \hline
{\it dog}  & 1.14E-36  & 5.56E-16 & 1.44E-46   & 3.34E-25 & 1.18E-32    & 3.03E-26 & 3.43E-7  \\ \hline
{\it hedgehog}  & 5.66E-17  & 8.07E-6 & 2.05E-7  & 0.16E-3 & 7.20E-5     & 0.14 & 3.48   \\ \hline
{\it guinea pig}  & 2.47E-7  & 5.67E-23 & 2.80E-24   & 7.57E-5 & 1.15E-5   & 1.60E-10 & 1.31  \\ \hline
 &  $e_3$/$1$ & $e_4$/$e_5$ & $e_4$/$e_6$ & $e_4$/$1$ & $e_5$/$e_6$ & $e_5$/$1$ & $e_6$/$1$\\ \hline   
{\it rabbit}  & 2.05E-19  & 3.14E-10 & 4.98E-11 & 0.12E-2 & 0.72E-1    & 6.40E-36 & 2.01E-39  \\ \hline
{\it cat}  & 5.20E-22  & 0.92 & 4.49E-5  & 3.49E-31 & 0.12E-3    & 8.79E-27 &  4.90E-48   \\ \hline
{\it mouse}  &  9/81E-7  & 2.00E-26 & 6.96E-27   & 3.28E-13 &  0.27   & 4.12E-22 &  7.22E-24  \\ \hline
{\it bird} & 8.08E-22  & 1.62E-12 & 9.37E-8   & 1.19E-6 &  2.02E-16   & 1.32E-7 &  1.36E-35  \\ \hline
{\it parrot}  & 5.14E-33  & 2.29E-34 &  1.89E-9   & 3.90E-7 &  4.03E-36   & 2.41E-33 &  1.63E-33   \\ \hline
{\it goldfish}  & 3.87E-49  & 8.69E-8 &  2.29E-58   &  2.35E-30 & 1.56E-58   & 1.85E-47 & 4.42E-54  \\ \hline
{\it hamster}  & 0.05  & 5.41E-10 & 6.17E-12    & 2.40E-12 & 0.26E-1   & 2.08E-37 & 5.90E-42   \\ \hline
{\it canary}  &  1.15E-41  & 0.16E-3 & 0.21E-3   & 2.15E-24 & 7.69E-7   & 0.56E-1 &  5.61E-32   \\ \hline
{\it guppy}  & 2.65E-42  & 4.24E-7 & 2.26E-53   & 2.64E-24 & 2.38E-54    & 8.60E-41 & 4.61E-51  \\ \hline
{\it snake}  & 6.19E-9  & 4.47E-32 &  6.46E-32    & 1.16E-28 & 0.39   & 1.53E-19 & 1.16E-9   \\ \hline
{\it spider}  & 0.15  & 1.56E-33 & 1.27E-33  & 3.23E-31 & 0.71E-1    & 3.07E-10 & 8.69E-11   \\ \hline
{\it dog}  & 4.50E-23  & 0.25E-2 & 1.71E-6   & 3.83E-30 & 7.17E-7    & 1.65E-6 & 1.09E-50  \\ \hline
{\it hedgehog}  & 4.22E-13  & 1.63E-22 & 3.80E-23  & 1.93E-16 & 0.84E-1     & 2.24E-13 & 5.04E-14   \\ \hline
{\it guinea pig}  & 0.72E-3  & 4.49E-10 & 9.28E-12   & 2.69E-9 & 0.52E-1   & 1.62E-29 & 1.16E-38  \\ \hline
\end{tabular}
\end{center}
\caption{p-values for the different exemplars and pairs of contexts} \label{tab:pvaluesexemplars}
\end{table}

\section{The Basic Structure of SCOP}
A SCOP consists not just of the three sets
$\Sigma, {\cal M}$, and ${\cal L}$: the set of states, the set of contexts and the set of
properties, but contains two additional functions $\mu$ and 
$\nu$. The function $\mu$ is a 
probability function that describes how state $p$ under the influence of context $e$ changes 
to state $q$. Mathematically, this means that $\mu$ is a 
function from the set $\Sigma \times {\cal M} \times \Sigma $ to the interval $[0, 
1]$, where $\mu(q, e, p)$ is the probability that state $p$ under the influence of context $e$ 
changes to state $q$. We write $\mu: \Sigma \times {\cal M} \times \Sigma \rightarrow [0, 1]; (q, e, p) \mapsto \mu(q, e, p)$. The function $\nu$ describes the weight (the renormalization of the
applicability) of a certain property given a specific state. This means that 
$\nu$ is a function from the set $\Sigma \times {\cal L}$ to the interval $[0, 1]$, where 
$\nu(p, a)$ is the weight of property $a$ for the concept in state $p$. We write $\nu: \Sigma \times {\cal L} \rightarrow [0, 1]; (p, a) \mapsto \nu(p, a)$.
Thus the 
SCOP is defined by the five elements $(\Sigma, {\cal M}, {\cal L}, \mu, \nu)$. Up until this point, the SCOP we have built for the concept `pet' has been rather small. To build a more elaborate SCOP, we proceed as follows. We collect all the contexts thought to be relevant to the model we want to build (more contexts lead to a more refined model). ${\cal M}$ is the set of these contexts. Starting from the ground state $\hat{p}$ for the concept, we collect all the new states of the
concept formed by having each context $e \in {\cal M}$ work on $\hat{p}$ and consecutively on all the
other states. This gives the set
$\Sigma$. Note that
${\cal M}$ and $\Sigma$ are connected in the sense that to complete the model it is necessary to consider the effect of each
context on each state. We collect the set of relevant properties of the concept and this gives ${\cal
L}$. The functions $\mu$ and $\nu$ that define the metric structure of the SCOP have to be determined by means of well chosen
experiments. First, however, we derive the natural structures that exist on the sets $\Sigma$, ${\cal M}$ and ${\cal L}$.

\subsection{The Lattice of Contexts} \label{sec:contextlattice}
By deriving a lattice structure for the set of contexts ${\cal M}$ we will be able to show that the set of contexts has a nonclassical (quantum-like) structure. First we identify a partial order relation on ${\cal M}$. Consider
for the concept `pet' the contexts
\begin{eqnarray}
&&e_3: {\rm `The\ pet\ runs\ through\ the\ garden'} \\
&&e_7: {\rm `The\ pet\ runs\ through\ the\ garden\ trying\ to\ catch\ a\ cat'} \\
&&e_8: {\rm `The\ pet\ runs\ through\ the\ garden\ trying\ to\ catch\ a\ cat\ while\ barking\ loudly'} 
\end{eqnarray}
These three contexts are related. We say that $e_7$ `is
stronger than or equal to'
$e_3$ and $e_8$ `is stronger than or equal to' $e_7$. Let us denote the relation `is stronger than or equal to' with the symbol $\le$. This means 
we have
$e_8
\le e_7
\le e_3$. It is easy to verify that the relation `is stronger than or equal to' is a partial order relation. This means that it satisfies the following
mathematical rules. For arbitrary contexts $e, f, g
\in {\cal M}$ we have
\begin{eqnarray}
{\rm reflexivity}: \quad && e \le e \label{eq:reflexivity} \\
{\rm transitivity}: \quad && e \le f,\ f \le g\ \Rightarrow\ e \le g \label{eq:transitivity} \\
{\rm symmetry}: \quad && e \le f, \ f \le e\ \Rightarrow\ e = f \label{eq:symmetry}
\end{eqnarray}
(\ref{eq:reflexivity}) means that each context is `stronger than or equal to' itself.
(\ref{eq:transitivity}) means that if a first context is `stronger than or equal to' a second, and this second is `stronger than or equal to' a third,
then the first is `stronger than or equal to' the third. (\ref{eq:symmetry}) means that if a first context is `stronger than or equal to' a second, and
this second is `stronger than or equal to' the first, then they are equal.

If a set is equipped with a partial order relation, it is always possible to verify whether for a subset of elements of the set, there exists an
infimum or greatest lower bound and a supremum or least upper bound of this subset with respect to this partial order relation. Hence,
consider ${\cal M}$, now equipped with the partial order relation
$\le$, and a subset $\{e_i\}_{i \in I}$ of elements $e_i \in {\cal M},\ \forall i \in I$. An element $\wedge_{i \in I}e_i \in {\cal M}$ is an
infimum of the subset $\{e_i\}_{i \in I}$ if it is a lower bound, which means that
$\wedge_{i \in I} e_i \le e_j\ \forall j \in I$
and additionally it is the greatest lower bound, {\it i.e.} a maximum of the set of all lower bounds of $\{e_i\}_{i \in I}$. This means that for
each possible context $f \in {\cal M}$ that is a lower bound---hence $f$ is such that $f \le e_j\ \forall j \in I$---we have that
$f
\le \wedge_{i \in I} e_i$. This expresses that $\wedge_{i \in I} e_i$ is the greatest lower bound (if a greatest lower bound exists it is always unique, which means that we can talk of `the' greatest lower bound).

Let us see what this somewhat subtle notion of infimum means with respect to the set of contexts for the concept `pet'. Consider the context 
\begin{equation}
e_9: {\rm `The\ pet\ tries\ to\ catch\ a\ cat'}
\end{equation}
We clearly have $e_7 \le e_9$. We already remarked that
$e_7 \le e_3$, which means that $e_7$ is a lower bound for $e_3$ and $e_9$. If we try to conceive of a context that is stronger
than or equal to $e_3$, `The pet runs through the garden', and also stronger than or equal to $e_9$, `The pet tries to catch a cat', it will also
be stronger than or equal to $e_7$, `The pet runs through the garden trying to catch a cat'. An example of such a context is
$e_8$, `The
pet runs though the garden trying to catch a cat while barking loudly'. This shows that $e_7$ is the infimum of $e_9$ and $e_3$, hence $e_7 =
e_9
\wedge e_3$.

This all shows that it is plausible to require that for an arbitrary subset of contexts $\{e_i\}_{i \in I}$, $e_i \in {\cal M},\ \forall i \in I$
there exists an infimum context $\wedge_{i \in I} e_i \in {\cal M}$. Mathematically we formulate this requirement as follows: For $\{e_i\}_{i
\in I}, e_i \in {\cal M},\ \forall i \in I$ there exists $\wedge_{i \in I}e_i \in {\cal M}$ such that for
$f \in {\cal M}$ we have
\begin{eqnarray}
{\rm lower\ bound}: \quad && \wedge_{i \in I}e_i \le e_j\ \forall j \in I \\
{\rm greatest\ lower\ bound}: \quad && f \le e_j\ \forall j \in I\ \Rightarrow f \le \wedge_{i \in I}e_i
\end{eqnarray}
Note that the infimum context corresponds to what could be called an `and' context, which is why we
denote it using the logical symbol $\wedge$ for `and'.

Now that the infimum or `and' context has been introduced, one can ask whether a supremum context exists. Consider the following
additional context for the concept `pet' 
\begin{equation}
e_{10}: {\rm `The\ pet\ runs\ through\ the\ garden\ barking\ loudly\ to\ be\ fed'}
\end{equation}
One can consider the sentence `The pet runs through the garden trying to catch a cat {\it or} barking loudly to be fed'. At first glance this sentence does not seem to describe a possible context. However we should not fall into the trap of
identifying the context that describes this sentence with one of its instances. It is difficult to conceive of an instance of what is expressed by a sentence like `The pet runs through the garden trying to catch a cat {\it or} barking loudly to be fed'. Indeed, the only thing one can think of is that it is an instance of `The pet runs through the garden trying to catch a cat' {\it or} an instance of `The pet runs through the garden barking loudly to be
fed'. This is because it is not possible to conceive of an instance that is typical for `The pet runs through the garden trying to
catch a cat {\it or} barking loudly to be fed' is that if we consider two instances---let us call them `instance 1' and `instance
2'---then something like `instance 1 {\it or} instance 2' is {\it not} and instance. The reason for this is deep and rooted in the
nature of the structure of the world. It is similar to the fact that `physical entity 1 {\it or} physical entity 2' is not a physical entity, and
`situation 1 {\it or} situation 2' is not a situation. Concretely, `an orange {\it or} a chair' is not an entity. It was noted in \cite{kamppartee01} that the prototype theory has difficulty representing the `or' concept, {\it e.g}, what is the prototype of `a butterfly {\it or} a vegetable'. Much as there do not exist typical instances of the concept `a chair {\it or} a vegetable', there does not exist a prototype of this concept.

One can now ask whether `The pet runs through the garden trying to catch a cat {\it or} barking loudly to be fed' is a context? In our approach it is, but to make this clear we must explain what this context is. We will call it
the {\it superposition context} of the context `The pet runs through the garden trying to catch a cat' and the context `The pet runs through the
garden barking loudly to be fed'. Let us define formally what a superposition context is. Suppose we have a subset of contexts $\{e_i\}_{i \in I}, e_i
\in {\cal M},\ \forall i
\in I$. The superposition context, denoted
$\vee_{i \in I} e_i$, consists of one of the contexts $e_i$ but we do not know which one. This means that if we introduce
explicitly
\begin{equation}
e_{11}: {\rm `The\ pet\ runs\ through\ the\ garden\ trying\ to\ catch\ a\ cat\ {\it or}\ barking\ loudly\ to\ be\ fed'}
\end{equation}
then we have $e_{11} = e_7 \vee e_{10}$. This superposition context $\vee_{i \in I} e_i$ is the supremum context for a subset of contexts $\{e_i\}_{i
\in I}, e_i \in {\cal M},\ \forall i \in I$. It is obviously an upper bound, and it is easy to verify that it is the least upper bound.

The infimum context $\wedge_{i \in I} e_i \in {\cal M}$ of a subset of contexts $\{e_i\}_{i
\in I}, e_i \in {\cal M},\ \forall i \in I$ is a context that is more concrete than the
two original contexts, and it can be expressed by the `and' of language if we express the contexts using sentences. The supremum context
$\vee_{i \in I} e_i \in {\cal M}$ of a subset of contexts $\{e_i\}_{i
\in I}, e_i \in {\cal M},\ \forall i \in I$ is a context that is more abstract than the two original contexts,
and it can be expressed by the `or' of language if we express the context using sentences.

A partially ordered set that has an infimum and a supremum for any subset of its elements is called a {\it complete lattice}. The
`complete' refers to the fact that the supremum and infimum exists for any subset. If they exist only for finite subsets and not necessarily for
infinite subsets the structure is called a lattice. We call
${\cal M},
\le $ equipped with the partial order relation `is stronger than or equal to' the lattice of contexts. 

\subsection{The Identification of Quantum Structure} \label{sec:quantumstructure}
This section, like the previous one, concerns the complete lattice structure of ${\cal M}$, but here it is approached somewhat differently.
Suppose we have a concept
$S$, described by a SCOP $(\Sigma, {\cal M}, {\cal L}, \mu, \nu)$. We say that $p \in \Sigma$ is an eigenstate of the context $e \in {\cal M}$ if
the state $p$ of a concept does not change when context $e$ is applied to it. Thus $p$ is an eigenstate of the concept for the context $e$ iff $\mu(p, e,
p) = 1$. For example, the state $p_3$ of the concept `pet' in the situation `The pet runs through the garden' is an eigenstate of the context $e_3$, `The
pet runs through the garden'. The state $p_{10}$ of the concept `pet' in the situation `The pet runs through the garden
barking loudly to be fed' is an eigenstate of the context $e_{10}$, `The pet runs through the garden barking loudly to be fed', but is also an eigenstate
of the context $e_3$, `The pet runs through the garden', and of the context 
\begin{equation}
e_{12}: {\rm `The\ pet\ barks\ loudly\ to\ be\ fed'}
\end{equation}
We hypothesize that if an arbitrary context $e \in {\cal M}$ changes an arbitrary state $p \in \Sigma$ to a state $q \in \Sigma$, then $q$ is an eigenstate of $e$. This amounts to requesting that if a context affects the concept again right after it has affected the concept a first time, this does not introduce an additional change. Such measurement contexts are called in quantum mechanics `measurements of the first kind'. Another way of stating our hypothesis is to say that we confine ourselves to contexts of the first kind, namely those that do not re-affect the state of the concept if re-applied immediately after their first application. In quantum mechanics there do exist measurement that are not of the first kind, and certainly in cognition there also exist contexts that are not of the first kind. Exactly as in quantum mechanics, the mathematical model is built for SCOP by considering the contexts of the first kind and afterward treating contexts that are not of the first kind as derived notions. A state that is not an eigenstate of a context is called a potentiality state with respect to this context. The effect of a context is to change a potentiality state of this context to an eigenstate of this context, and this change will be referred to as collapse. The complete lattice structure introduced in Section
\ref{sec:contextlattice} can now be re-introduced in a slightly different way. Let us define the map 
\begin{eqnarray} \label{eq:lambda}
\lambda: {\cal M} &\rightarrow& {\cal P}(\Sigma) \\
e &\mapsto& \lambda(e) \\
\lambda(e) &=& \{p\ \vert\ p\ {\rm eigenstate\ of}\ e\} = \{p\ \vert\ \mu(p, e, p) = 1\}
\end{eqnarray}
where ${\cal P}(\Sigma)$ is the set of all subsets of $\Sigma$. We then define for $e, f \in {\cal M}$
\begin{equation}
e \le f \Leftrightarrow \lambda(e) \subset \lambda(f)
\end{equation}
which is the partial order relation `stronger than or equal to' considered in Section \ref{sec:contextlattice}. We suppose that for any
subset of contexts
$\{e_i\}_{i
\in I}, e_i \in {\cal M},\ \forall i \in I$ there exists an infimum context $\wedge_{i \in I} e_i$ and a supremum
context $\vee_{i \in I} e_i$, which makes ${\cal M}$ into a complete lattice. We denote the zero context $\wedge_{e \in {\cal M}} e$ by $0$. It
is the context that has no eigenstates. We denote the unit context $\vee_{e \in {\cal M}} e$ by $1$. It is the context for which each state is an
eigenstate.

For the infimum context $\wedge_{i \in I} e_i$ we have
\begin{eqnarray}
\cap_{i \in I}\lambda(e_i) &=& \lambda(\wedge_{i \in I} e_i)
\end{eqnarray}
This means that $p \in \Sigma$ is an eigenstate of $\wedge_{i \in I} e_i$ iff $p$ is an eigenstate of each of the $e_i$, which is why we can call it the
`and' context. However, for the supremum context, we do not have the equivalent equality. We have only
\begin{equation} \label{eq:quantum}
\cup_{i \in I}\lambda(e_i) \subset \lambda(\vee_{i \in I} e_i)
\end{equation}
and in general not the equality of these two expressions. The fact that we do not have an equality here is what makes our structure quantum-like.

Let us illustrate this with an example. Consider context $e_{11} = e_7 \vee e_{10}$. $\lambda(e_7)$ is the set of eigenstates
of `pet' for the context $e_7$, `The pet runs through the garden trying to catch a cat', and $\lambda(e_{10})$ is the set of eigenstates of
`pet' for the context $e_{10}$, `The pet runs through the garden barking loudly to be fed'. The set $\lambda(e_7) \cup \lambda(e_{10})$ is
the union of the two sets $\lambda(e_7)$ and $\lambda(e_{10})$. This means that if $p \in \lambda(e_7) \cup
\lambda(e_{10})$ we must have $p \in \lambda(e_7)$ {\it or} $p \in \lambda(e_{10})$. Hence $p$ is an eigenstate of the context $e_7$ {\it or} $p$
is an eigenstate of the context $e_{10}$. Now consider the state of `pet' corresponding to the situation `The pet runs through the garden trying
to catch a cat {\it or} barking loudly to be fed'. We do not know which of the two alternatives, `trying to catch a cat' or
`barking loudly to be fed' it is, but we know it is one of them. Let us describe a concrete situation where 
`pet' is in this state. Suppose one has a pet that only runs through the garden barking loudly in two situations: when it is
trying to catch a cat, and when it was hungry and wants to be fed. We call home and learn that the pet is running through the garden barking loudly. So
we do not hear the barking ourselves, because this would provide enough information for us to know which one of the two it is (perhaps the pet barks differently
in the two situations). So that is the state of the concept `pet' for us at that moment. Let us call this
state $p_{11}$. Then $p_{11}$ is an eigenstate of the context $e_{11} = e_7 \vee e_{10}$. Hence this means that $p_{11} \in \lambda(p_7 \vee
p_{10})$. But $p_{11}$ is not an eigenstate of $e_7$ and it is not an eigenstate of $e_{10}$. Indeed, if $p_{11}$ were an eigenstate of
$e_7$, this would mean that in state $p_{11}$ the pet would be running through the garden trying to catch a cat, and if it were an
eigenstate of $e_{10}$ this would mean that the pet was running through the garden barking loudly to be fed. Neither is true, which shows
that $p_{11} \not\in \lambda(e_7) \cup \lambda(e_{10})$. Thus $\lambda(e_7) \cup \lambda(e_{10}) \not= \lambda(e_7 \vee e_{10})$,
which proves that (\ref{eq:quantum}) is a strict inclusion and not an equality. It turns out that $p_{11}$ is a superposition state when we
represent the SCOP in the Hilbert space of quantum mechanics \cite{aertsgabora01}.

Applying analogous techniques to those in
\cite{aertsetal08} and \cite{aertsetal09}, it can be proven that $\lambda({\cal M})$ is a closure space. Further structural results about
SCOP can be obtained along the lines of
\cite{aertsetal10,aertsdeses01,aertsetal11,aertsdeses02}. Having introduced this closure space
the supremum can be given a topological meaning, namely for $e, f
\in {\cal M}$ we have $\lambda(e \vee f) = \overline{\lambda(e) \cup \lambda(f)}$
where $\overline{\lambda(e) \cup \lambda(f)}$ is the closure of $\lambda(e) \cup \lambda(f)$, which is obtained exactly, in the case of the
linear Hilbert space introduced in \cite{aertsgabora01}, by adding the superposition states to the
set $\lambda(e) \cup \lambda(f)$. This is why $e \vee f$ was called the superposition context of $e$ and $f$.

\subsection{Orthocomplementation and Atomicity} \label{sec:orthocomplementationatomicity}
There is another structure on ${\cal M}$ that is as important as the complete lattice structure. Consider the context
$e_3^\perp$, `The
pet does not run through the garden'. This context has a special relation to the context
$e_3$, `The
pet runs through the garden'. We call $e_3^\perp$ the {\it orthocomplement} of context $e_3$. Obviously we have $(e_3^\perp)^\perp = e_3$. Consider the
context $e_3$, `The pet runs through the garden', and the context $e_7$, `The pet runs through the garden trying to catch a cat'. Then we have $e_7
\le e_3$. The context $e_7^\perp$ is the following: `The pet does not run through the garden trying to catch a cat'. Hence we have $e_3^\perp \le
e_7^\perp$. We know that $e_7 = e_3 \wedge e_9$. The context $e_7^\perp$ can also be expressed as follows: `The
pet does not run through the garden {\it or} does not try to catch a cat'. This means that $e_7^\perp = e_3^\perp \vee e_9^\perp$. The orthocomplement is characterized mathematically as follows: $^\perp$ is a map from ${\cal M}$ to ${\cal M}$ such that for $e, f \in {\cal M}$ we have
\begin{eqnarray}
&&(e^\perp)^\perp = e \\
&&e \le f \Rightarrow f^\perp \le e^\perp \\
&&e \wedge e^\perp = 0, e \vee e^\perp = 1
\end{eqnarray}
The set ${\cal M}, \le, ^\perp$ is a complete orthocomplemented lattice. It can easily be seen that the orthocomplement is not a complement, due to the
existence of superposition states. For example, the ground state $\hat{p}$ of the concept `pet' is neither an eigenstate of the context $e_3$, `The pet
runs through the garden', nor of the context $e_3^\perp$, `The pet does not run through the garden'. This means that although $\lambda(e_3) \cup
\lambda(e_3^\perp) \subset \lambda(e_3 \vee e_3^\perp)$, this is a strict inclusion, hence
$\lambda(e_3) \cup \lambda(e_3^\perp) \not= \lambda(e_3 \vee e_3^\perp)$, which would not be the case if the orthocomplement were a
complement. Considering the closure space introduced by $\lambda$ we do have $\lambda(e_3 \vee e_3^\perp) = \overline{\lambda(e_3) \cup \lambda(e_3^\perp)}$.
It also means that the underlying logic is not classical but paracomplete \cite{aertsetal02}. In \cite{aertsgabora01} it turns out that the states that are in $\overline{\lambda(e_3) \cup \lambda(e_3^\perp)}$ and not contained in $\lambda(e_3) \cup \lambda(e_3^\perp)$, are the superpositions of
states in $\lambda(e_3) \cup \lambda(e_3^\perp)$.

Another important notion is that of the atom of a lattice. Consider a concept and a context $c \in {\cal M}$ such that $c \not=0$, and for $a \in {\cal M}$ we have that $a \le c$ implies that
$a = 0$ or $a = c$, then $c$ is called an atomic context of the concept. An atomic context is a strongest context different
from the zero context. 

To make this clearer, consider for a moment a small SCOP $(\Sigma, {\cal M}, {\cal L}, \mu, \nu)$ of the concept
`pet' containing the contexts $e_1, e_2, e_6$ as they appear in Table \ref{tab:differentcontexts}. The zero context and the 1 context are also 
elements of
${\cal M}$. For each context, the orthocomplement of this context is also an element of ${\cal M}$. This gives already $\{0, 1, e_1, e_1^\perp,
e_2, e_2^\perp, e_6, e_6^\perp\} \subset {\cal M}$. Furthermore, we need to add all the infima
different from the zero context and all the suprema different from the unit context. For example $e_1 \wedge e_2$ is the context `The pet is
chewing a bone and being taught', and this is not the zero context. The context $e_1
\wedge e_6$ would probably normally be classified as equal to the zero context. Indeed, a pet that is a fish does not chew a bone. However, we
can invent situations where even this infimum would not be equal to the zero context. Consider for example a movie in which a child dreams of getting a dog as a pet, but receives a fish. In the movie, the fish is a conscious and intelligent being, and knows the desire of the
child, and just to make her happy decides to behave as much as possible like a dog. Hence chewing a bone is something that is tried out by
the fish. The example shows that the situation would be possible where another child leaving the movie theatre with her mother, says: ``Mom, that was too funny how the fish was chewing a bone". Since we do not want to exclude from our theory of
concepts these more exotic situations, we do not have to put $e_1 \wedge e_6 = 0$ a priori. However, what is really happening here is
that different SCOPs can be constructed starting with the same set of contexts, because one needs to decide whether some
infima will be equal to the zero context or not. But this is all right. Each different SCOP is another model. If we want to construct
a model of the concept `pet' where the possibility that the pet is a fish that is chewing a bone is important, we need to allow $e_1 \wedge
e_6$ to be different from zero. If we are not interested in this, and want a simpler model, we can put $e_1 \wedge e_6 = 0$. 

Because it is
our aim to show what an atomic context is, we will choose a simple model. So we put $e_1 \wedge e_6 = 0$. Having tried out
all the possible infima and suprema for a simple case we get the following: ${\cal M} = \{0, 1, e_1, e_1^\perp, e_2, e_2^\perp, e_6, e_6^\perp,
e_1
\wedge e_2, e_1
\wedge e_2^\perp, e_1 \wedge e_6^\perp,
e_1^\perp \wedge e_2, e_1^\perp \wedge e_2^\perp, e_1^\perp \wedge e_6, e_1^\perp \wedge e_6^\perp, e_2 \wedge e_6^\perp, e_2^\perp \wedge e_6,
e_2^\perp
\wedge e_6^\perp,
e_1^\perp \vee e_2^\perp, e_1^\perp
\vee e_2, e_1^\perp \vee e_6,
e_1 \vee e_2^\perp, e_1 \vee e_2, e_1 \vee e_6^\perp, e_1 \vee e_6, e_2^\perp \vee e_6, e_2 \vee e_6^\perp,
e_2
\vee e_6 \}$. The set of atomic context, denoted ${\cal A}({\cal M})$, is given by: ${\cal A}({\cal M}) = \{e_1 \wedge e_2, e_1
\wedge e_2^\perp, e_1 \wedge e_6^\perp,
e_1^\perp \wedge e_2, e_1^\perp \wedge e_2^\perp, e_1^\perp \wedge e_6, e_1^\perp \wedge e_6^\perp, e_2 \wedge e_6^\perp, e_2^\perp \wedge e_6,
e_2^\perp
\wedge e_6^\perp \}$, counting 10 elements.

Given a context $e \in {\cal M}$ of a concept $S$, sometimes we have called $p$ the state corresponding to this context. When we write this, we
mean the state $p$ that the context is in under the effect of the context $e$ if it was in the ground state $\hat{p}$ before the context $e$
started to influence it. If the concept is in another state than the ground state, the same context $e$ will influence the concept such that it
is in general transferred to another state than $p$. This means that there is no unique state that corresponds to a specific context. How to connect a set of states (and hence not a unique state) to a context is expressed by the function
$\lambda$ as defined in (\ref{eq:lambda}). For a context $e \in {\cal M}$ the set $\lambda(e)$ is the set of eigenstates of $e$. The state that
we have denoted $p$, and that is the state that the context $e$ changes the ground state $\hat{p}$ to, is one of these eigenstates.

\subsection{The Orthocomplemented Lattice of Properties} \label{sec:propertylattice}
A property of a concept is described using the notions of actuality, potentiality, and weight of the property as it relates to the
state of the concept. That is how gradedness is accounted for. Referring back to Tables \ref{tab:properties} and 
\ref{tab:applicabilityratings}, consider the concept `pet' in its ground state
$\hat{p}$. The property $a_1$, {\it lives in and around the house} received a very high rating for the ground state. Subjects estimated that
this is a property that is almost always actual for the ground state of the concept `pet'. Only under context $e_4$, `Did you see the type of
pet he has? This explains that he is a weird person', did the weight of property $a_4$ substantially decreases. A `weird pet' is considered by
the subjects to live less in and around the house. But property $a_1$ can be considered as a very characteristic property for the concept `pet',
which means that moving across the different states of `pet' there is not much change of actuality to potentiality and vice versa. This is not the case for property
$a_6$, {\it can fly}. Subjects rated this property with weight 0.57 in the ground state, which means that the property is considered to be actual around
half of the time and potential the other half for `pet' in the ground state. However for `pet' in state $p_1$---hence under context
$e_1$---`The pet is chewing a bone', the rating decreases to 0.14, while for pet in state $p_5$, hence under context $e_5$, `The pet is
being taught to talk' it increases to 0.86. Table \ref{tab:applicabilityratings} shows how a property changes
weight when the concept `pet' changes from state to state under different contexts.

We say that a property $a \in {\cal L}$ of a concept $S$ is {\it actual} in state $p \in \Sigma$ iff $\nu(p, a) = 1$. This makes
it possible to introduce a partial order relation on ${\cal L}$ as follows. For $a, b \in {\cal L}$ we have
\begin{equation}
a \le b \Leftrightarrow a\ {\rm is\ actual\ in\ state}\ p\ {\rm then}\ b\ {\rm is\ actual\ in\ state}\ p 
\end{equation}
We say that $a$ is `stronger or equal to' $b$. This makes ${\cal L}$ into a partially ordered set. Hence we can ask about the existence
of an infimum and a supremum for this partial order. Consider a subset $\{a_i\}_{i \in I}$ of properties $a_i \in {\cal L},\ \forall i \in I$ 
of the concept $S$. We denote the infimum property by $\wedge_{i \in I} a_i$ and the supremum property by $\vee_{i \in I} a_i$. It can be shown
that the infimum property is a `and' property and the supremum property is a `or' property. We have, for $a_i \in {\cal L}, i \in I$
\begin{eqnarray}
a_i\ {\rm actual\ for\ all}\ i \in I  &\Leftrightarrow& \wedge_{i \in I} a_i\ {\rm actual} \label{eq:infimum}\\
{\rm one\ of\ the}\ a_j\ {\rm actual\ for}\ j \in I &\Rightarrow& \vee_{i \in I} a_i\ {\rm actual} \label{eq:supremum}
\end{eqnarray}
It is important to remark that for (\ref{eq:supremum}) the implication $\Leftarrow$ is in general not true. Indeed, consider the
concept `pet' and the properties $a_{14}$, {\it friendly}, and $a_{15}$, {\it not friendly}. Then for an arbitrary state of `pet' we have that $a_{14}
\vee a_{15}$ is actual, {\it i.e.} `the pet is {\it friendly} or is {\it not friendly}'. But this does not mean that for this same state $a_{14}$ is actual
{\it or}
$a_{15}$ is actual. Indeed, it is common to encounter a state $p$ of `pet' where $\nu(p, a_{14}) \not= 1$ and
$\nu(p, a_{15}) \not= 1$. For example, the ground state $\hat{p}$ is like this, as are the states $p_1, p_2, \ldots, p_6$.

The structure of the set of properties of a physical entity has been the subject of intense research in quantum axiomatics
\cite{aerts01,aerts02,aerts03,aerts04,aerts05,aerts06,aerts07,foulisetal01,foulisrandall01,jauch01,piron01,piron02,piron03,pitowsky01,randallfoulis01,randallfoulis02,randallfoulis03,randallfoulis04}.
Most of the results obtained in quantum axiomatics can be applied readily to concepts being considered as entities with properties. Borrowing
from the study of State Property Systems
\cite{aerts05,aerts06,aertsetal08,aertsetal09,aertsetal10,aertsdeses01,aertsdeses02} we introduce the function
\begin{eqnarray}
\kappa: {\cal L} &\rightarrow& {\cal P}(\Sigma) \\
a &\mapsto& \kappa(a) \\
\kappa(a) &=& \{p\ \vert\ p\ {\rm makes\ the\ property}\ a\ {\rm actual}\}
\end{eqnarray}
that has been called the `Cartan Map' in the study of State Property Systems. Clearly we have for $a \in {\cal L}$
\begin{equation}
\kappa(a) = \{q\ \vert\ \nu(q, a) = 1, q \in \Sigma\}
\end{equation}
and for $a, b \in {\cal L}$
\begin{equation}
a \le b \Leftrightarrow \kappa(a) \subset \kappa(b)
\end{equation}
The Cartan Map introduces a closure space, namely $\kappa({\cal L})$.
By means of the closure space
the supremum can be given a topological meaning, namely for $a, b
\in {\cal L}$ we have
\begin{equation}
\kappa(a \vee b) = \overline{\kappa(a) \cup \kappa(b)}
\end{equation}
where $\overline{\kappa(a) \cup \kappa(b)}$ is the closure of $\kappa(a) \cup \kappa(b)$, which is obtained exactly, in the case of the
linear Hilbert space introduced in the paper that follows by adding the superposition states to
the set $\kappa(a) \cup \kappa(b)$.

The complete lattice of properties ${\cal L}$ also contains the natural structure of an orthocomplementation. Consider the property
$a_3^\perp$, {\it not feathered}. This property has a special relation to the property
$a_3$, {\it feathered}. We say that $a_3^\perp$ is the {\it orthocomplement} of property $a_3$. Obviously we have $(a_3^\perp)^\perp = a_3$. 
Now consider the property $a_{15}$, {\it feathered and can swim}. Then we have $a_{15} \le a_3$. The property $a_{15}^\perp$ is the
following: {\it not feathered and able to swim}. Hence we have $a_3^\perp \le a_{15}^\perp$. Now recall the property, {\it able to swim}, denoted $a_7$. We
know that $a_{15} = a_3 \wedge a_7$. And the property $a_{15}^\perp$ can also be expressed as follows: {\it not feathered or unable to swim}. This means
that $a_{15}^\perp = a_3^\perp \vee a_7^\perp$. Once again, the orthocomplementation can be characterized mathematically as follows: $^\perp$ is a map from ${\cal L}$ to ${\cal L}$ such that for $a, b \in {\cal L}$ we
have
\begin{eqnarray}
&&(a^\perp)^\perp = a \\
&&a \le b \Rightarrow b^\perp \le a^\perp \\
&&a \wedge a^\perp = 0, a \vee a^\perp = 1
\end{eqnarray}
The set ${\cal L}, \le, ^\perp$ is a complete orthocomplemented lattice. It can easily be seen that the orthocomplement is not a complement, due to
the existence of superposition states. For example, the ground state $\hat{p}$ of the concept `pet' is neither an eigenstate of the property $a_3$,
{\it feathered}, nor of the property $a_3^\perp$, {\it not feathered}. This means that although $\kappa(a_3) \cup \kappa(a_3^\perp) \subset \kappa(a_3
\vee a_3^\perp)$, this is a strict inclusion. Hence
$\kappa(a_3) \cup \kappa(a_3^\perp) \not= \kappa(a_3 \vee a_3^\perp)$, which would be the case if the orthocomplement were a complement.
Considering the closure space introduced by $\kappa$ we do have $\kappa(a_3 \vee a_3^\perp) = \overline{\kappa(a_3) \cup \kappa(a_3^\perp)}$. We will see that the states in
$\overline{\kappa(a_3) \cup \kappa(a_3^\perp)}$ and not contained in $\kappa(a_3) \cup \kappa(a_3^\perp)$, are the superpositions of
states in $\kappa(a_3) \cup \kappa(a_3^\perp)$.

\subsection{Concepts, Properties and Contexts}
In SCOP a concept is described by making use of the sets of contexts and properties relevant for this concept. However a property often is in itself a concept or an aggregation of concepts. One can wonder whether this does not lead to circularity. We can get an insight into this question by considering the situation in quantum mechanics. One of the relevant physical quantities of a quantum entity is its position. The context corresponding to a position measurement is a detector screen. The detector screen in itself is a congregation of quantum entities, namely the atoms and molecules that are the building blocks of the screen. In quantum mechanics the detector screen is described by another mathematical notion than a quantum entity or a congregation of quantum entities. The states of a quantum entity are described by vectors in a Hilbert space while the detector screen is described by an orthogonal projection operator of the same Hilbert space. In \cite{aertsgabora01} the states of a concept are described by vectors of a Hilbert space and a context by an orthogonal projection operator of the same Hilbert space. This means that also concerning this question of circularity we are in an analogous situation. The reason that this does not lead to circularity is the following: if a specific physical entity (a specific concept) is the focus of description by quantum mechanics (by SCOP), then the symmetry is broken, because the detector screen (context) is not the focus of description. There could only eventually be a problem of consistency. If the detection screen (context) is described by means of procedure for the description of the compound of different quantum entities (concepts; this is the procedure that we develop explicitly in \cite{aertsgabora01}), then this more refined description should give the same results as the standard one by means of an orthogonal projection operator. In quantum mechanics this problem is known as the measurement problem. We plan to study the equivalent of the measurement problem of quantum mechanics for concepts in future research.

As for properties being also aggregations of concepts, the analogy with quantum systems is not there any longer. Properties of a quantum entity are not aggregations of quantum entities, but they are too aggregations of concepts. However the same more general argument applies. Since the focus is on one and only one concept (or on one and only one combination of concepts), a property may, without leading to a problem of circularity be described in another way than the concept is being done. Hence we can described a property by means of an orthogonal projection operator, as is done in \cite{aertsgabora01}. Of course a similar problem of consistency appears in the following case: suppose a property is explicitly used in a sentence, and we want to describe this sentence as a combination of concepts. In this case, also the property will be described as a concept on its own, with the focus on this one and only concept that in the sentence fulfills the role of a property.

 \section{Summary and Conclusions}
In {\it Notes on an Epistemology for Living Things}, von Foerster writes "Objects and events are not primitive experiences. Objects and events are representations of relations \cite{vonfoerster03}." This insightful comment shows clearly that at some level he foresaw the step we have taken in this paper. If we see something and classify it as an instance of `bird', we are forging a relationship between a context---this particular experience of a particular bird chirping in a particular tree---and our concept `bird'. How one experiences the concept `bird' depends on the circumstances that evoked it. 

Consider the concept `pet' in the two following contexts---`The pet is chewing a bone' and `The pet is being taught to talk'. If subjects are asked to rate the typicality of a specific exemplar of `pet' and the applicability of a particular property of `pet', their ratings will depend on whether `pet' is considered under the first context or the second. The exemplar {\it dog}, for example, will rate high under the first context and low under the second context, whereas the exemplar {\it parrot} will show the inverse . Similarly with properties, {\it furry} will rate high under the first context and low under the second, whereas {\it feathered} will show the inverse pattern. A basic aim of our formalism is to model this type of contextual influence on a concept. Our method of incorporating contextual influence enables us to model the combination of concepts, and hence proposes a solution to a problem that is considered to be very important and wholly unsolved within existent theories of concepts, {\it i.e.} the combination problem.

To incorporate the effect of contextual influence, our theory introduces the notion of `state of a concept'. For the example above we introduce two states of the concept `pet', {\it i.e.} one that accounts for the ratings under the first context, and another that accounts for the ratings under the second. Thus our theory considers a concept to be an entity comprising different states, with each of the states accounting for the different exemplar typicalities and property applicabilities. Note that we are not just proposing that the applicabilities of properties differ amongst different exemplars of a concept, an effect well accounted for in other theories, {\it e.g.} prototype and exemplar theories. The applicability of a single property varies for each state, as does the typicality of a single exemplar. 

Our theory proposes the structure of a State Context Property System (SCOP) to model a concept. SCOP consists of a set of relevant states $\Sigma$, a set of relevant contexts ${\cal M}$, and a set of relevant properties ${\cal L}$. An archetypical `change' modeled by the SCOP is the following. The concept, when in a specific state $p$ contained in $\Sigma$, accounting for the typicality values of exemplars and applicabilities of properties (in that state $p$), changes to another state $q$ contained in $\Sigma$ under a specific context $e$ contained in ${\cal M}$, in its turn accounting for the changed typicality values of exemplars and changed applicabilities of properties (in that state $q$). It is this possibility of `dynamic change' under the influence of a context within SCOP that allows us to model the combination of concepts. When concepts combine, they mutually affect how they function as a context for each other, and hence provoke the type of dynamical change of state that is a basic aspect of our theory. Existing theories of concepts are unable to describe combinations of concepts because they have no means to describe the dynamical change of state under the influence of a context, which means that they can neither describe the change of state that one concept causes to other concepts in a combination of concepts, where this one concept functions as a context for these other concepts.

Section \ref{sec:scop} develops the mathematical structure of the SCOP in its most general form by identifying structures in the set of contexts ${\cal M}$ and the set of properties ${\cal L}$. If we consider, for the concept `pet', contexts $e$ `The pet runs through the garden trying to catch a cat' and $f$ `The pet runs through the garden', we can say that $e$ `is stronger than or equal to' $f$, thereby introducing a partial order relation in the set of contexts ${\cal M}$. By introducing the `and' context and the `or' context of two contexts, set ${\cal M}$ can be shown to obtain the structure of a complete lattice (Section \ref{sec:contextlattice}). By introducing the `not' context for any other context, the structure of an orthocomplementation can be derived for ${\cal M}$ (Section \ref{sec:orthocomplementationatomicity}). We then introduced the notions of `eigenstate' and `potentiality state' for a context. A state of a concept is said to be an eigenstate for a context of this concept if the state is not affected by the context. If the state is not an eigenstate of a context, it is said to be a potentiality state for this context (Section \ref{sec:quantumstructure}).

The quantum-like structure of the SCOP is revealed if we consider that for two contexts $e$ and $f$ contained in ${\cal M}$, a state will generally not be an eigenstate of the context $e$ `or' $f$, if and only if it is an eigenstate of $e$ `or' an eigenstate of $f$ (Section \ref{sec:quantumstructure}). Similarly, this quantum-like structure of the SCOP is revealed if we consider that although any state is an eigenstate of the context $e$ `or' not $e$, we cannot say that any state is an eigenstate of $e$ `or' an eigenstate of not $e$. The latter argument can easily be illustrated by means of the contrast between context $e$ `The pet runs through the garden', and context not $e$ `The pet does not run through the garden'. Any state that does not tell us what the pet is doing is neither an eigenstate of $e$ (indeed, in $e$ the state of pet changes to one in which the pet `runs through the garden') nor is it an eigenstate of not $e$ (in context not $e$, the state of pet is also affected, for it changes to a state in which the pet `does not run through the garden') (Section \ref{sec:orthocomplementationatomicity}).

A similar structure, namely that of a complete orthocomplemented lattice, can be derived for the set of properties ${\cal L}$ of the SCOP. This structure too can be shown to be quantum-like (Section \ref{sec:propertylattice}). The existence of a complete lattice structure for the sets of contexts and properties makes it possible to construct a topological representation of a SCOP in a closure space. In this closure space, the potentiality states whose presence makes the SCOP quantum-like are recuperated by the closure operation
(Sections \ref{sec:quantumstructure} and \ref{sec:propertylattice})

The identification of the complete orthocomplemented lattice structure for the sets of contexts and properties of the SCOP is an operational derivation, {\it i.e.} we do not make any non-operational technical hypothesis, but merely derive the structure by taking into account the natural relations (such as the partial order relation of `stronger than or equal to') that exist in the sets of contexts and properties.  

\bigskip
\noindent
{\bf Acknowledgments:} We would like to thank Alex Riegler and six anonymous reviewers for comments on the manuscript. This research was supported by Grant G.0339.02 of the Flemish Fund for Scientific Research.

\end{document}